\documentclass[pra,aps,showpacs,superscriptaddress,twocolumn]{revtex4}
\usepackage{graphicx}
\usepackage{amsmath}
\usepackage{amsfonts}
\usepackage{amssymb}
\usepackage{epsfig}

\begin{document}

\title{
On the finite size behavior of quantum collective spin systems}
\author{Giuseppe Liberti}\email{liberti@fis.unical.it}\affiliation{
Dipartimento di Fisica, Universit\`a della Calabria, 87036
Arcavacata di Rende (CS) Italy}
\author{Franco Piperno}\affiliation{ Dipartimento di Fisica, Universit\`a della
Calabria, 87036 Arcavacata di Rende (CS) Italy} \affiliation{INFN
- Gruppo collegato di Cosenza, 87036 Arcavacata di Rende (CS)
Italy}
\author{Francesco Plastina}\affiliation{
Dipartimento di Fisica, Universit\`a della Calabria, 87036
Arcavacata di Rende (CS) Italy}\affiliation{INFN - Gruppo
collegato di Cosenza, 87036 Arcavacata di Rende (CS) Italy}
\date{\today}

\begin{abstract}
We discuss the finite size behavior of the adiabatic Dicke model,
describing the collective coupling of a set of $N$-two level atoms
(qubits) to a faster (electromagnetic) oscillator mode. The energy
eigen-states of this system are shown to be directly related to
those of another widely studied collective spin model, the
uniaxial one. By employing an approximate continuum approach, we
obtain a complete characterization of the properties of the
latter, which we then use to evaluate the scaling properties of
various observables for the original Dicke model near its quantum
phase transition.
\end{abstract}
\bigskip
 \maketitle

\section{Introduction}

The interaction of $N$ two-level systems (qubits) with a common
single-mode quantum bosonic field is a paradigmatic example of
collective quantum behavior. Dating back to the model put forward
by Dicke, \cite{dicke}, this has become one of the most
investigated problems in quantum optics and condensed matter
physics, with proposed physical implementations ranging from
superconducting nano-devices, \cite{chen0}, to ultracold atoms and
Bose-Einstein condensates in cavity \cite{larson}. The Dicke model
exhibits a second-order phase transition \cite{hepp}, and, due to
its broad application range \cite{phyrep}, it has been studied
extensively in the past few years,
\cite{WH,gilmore,liberti1,fidelity}. It displays a rich dynamics,
with many non-classical features
\cite{milburn,brandes,Hou,orszagent}; in particular, the ground
state entanglement \cite{lambert,lambert2} and the Berry phase
\cite{liberti2,chen1} of the Dicke Model have been diffusely
analyzed and many aspects of its finite size behavior have been
obtained \cite{vidal1,liberti3,chen}.

The continued interest in the Dicke model also stems from the fact
that it pertains to the same universality class as other intensely
studied many-body systems that possess infinite-range
interactions, and for which theoretical models typically allow for
exact solutions in the thermodynamic limit.

The most general collective model of (effective) spin $1/2$
systems, the biaxial model in arbitrary field, can be described by
the Hamiltonian (see Ref. \cite{vidalconcurrence} for details)
\begin{equation}\label{biaxial}
\hat{H}_{XY}^{\perp,\,\parallel}=\sum_{k=x,y,z}{\delta_k}{\hat{S}_k}
+ g_x\hat{S}_x^2+g_y\hat{S}_y^2
\end{equation}
where the $\hat{S}_k = \sum_{i=1}^N \hat{\sigma}_i^{(k)}$ are the
collective Pauli operators that obey angular momentum-like
commutation relations
$[{\hat{S}}_{i},{\hat{S}}_{j}]=2i\epsilon_{ijk}{\hat{S}}_{k}$. The
energy eigenstates can be written in the angular momentum basis
(we employ the standard one, apart from a factor $2$ in the
definitions) $\{|s,s_z\rangle;\, s_z=-s,-s+2,\dots,s-2,s\}$
constructed as the set of common eigenstates of both
$\hat{S}^2=\hat S_x^2+\hat S_y^2 + \hat S_z^2$ and $\hat{S}_z$.
For a ferromagnetic interaction $g_{x,y}<0$, the ground state of
the Hamiltonian belongs to the symmetric sub-space with
${S}^2=N(N+2)$ and special and diffusely studied cases are the
biaxial model in a transverse field ($\delta_x=\delta_y=0$)
$\hat{H}_{XY}^{\perp}$ (the well-known LGM model
\cite{lipkin,lmgnuovi}) and the uniaxial model ($\delta_y=g_y=0$)
$\hat{H}_{X}^{\perp,\,\parallel}$.

Under the thermodynamic limit, the phase diagram of these
collective spin models has been simply established by a mean field
approach \cite{botet}. For $N$ large but finite, purely quantum
effects become important and numerical analysis have been
implemented using the continuous unitary transformation method
\cite{dusuel} and a semiclassical approach \cite{cui}. A
qualitative understanding of the LGM-model is obtained in
Ref.\cite{Heiss} introducing a double well structure above the
phase transition in a semiclassical treatment of the system.

In the present work, we establish an exact relationship between
the Dicke model in the adiabatic regime (i.e. for the case of slow
qubits coupled to a faster oscillator mode) and the uniaxial
model, which is valid not only in the thermodynamic limit, but
also for any finite number $N$ of spins. We then present an
alternative analytic method which relies on a continuum approach
to solve the collective uniaxial model for large $N$. We show that
this method is useful to determine the finite size behavior and
the entire $1/N$ expansion (i.e., critical exponents and
pre-factors) at the critical point for both of the  Dicke and the
collective uniaxial spin models. These results corroborate several
studies in which the exponents have already been derived.

The objective of the present study is thus threefold, and the
paper is organized accordingly: first, we consider the Dicke model
in the regime in which the frequency of the quantum field is much
larger than the energy spacing of the qubits; in this case, the
field degree of freedom can be adiabatically separated from the
qubit ones and an effective $N$-qubit interaction can be obtained
by means of the Born-Oppenheimer approximation. This is done in
Sec.\ref{sect2}, where the relationship with the uniaxial model is
established for any energy eigen-state. Afterwards, we focus on
the quantum phase transition of this collective model (Sec.
\ref{sect3}) for which we derive the $1/N$ expansion for some
relevant physical observables and we also compute exactly various
entanglement measures for the qubits. Finally, using these results
together with those obtained in Sec. \ref{sect2}, we obtain
analogous $1/N$ expansions for the Dicke model (Sec. \ref{sect4}).
A summary and some concluding remarks are finally given in Sec.
\ref{sect5}.

\section{Adiabatic Dicke model}\label{sect2}

We consider a system of $N$ qubits interacting with a single
harmonic oscillator mode, described by the Hamiltonian
($\hbar=c=1$)
\begin{equation}
\hat{H}=-\frac{\delta}{2} \hat{S}_x+\frac{\epsilon}{2} \hat{S}_z+
\omega \hat{a}^{\dagger}\hat{a} +
\frac{\lambda}{\sqrt{N}}(\hat{a}^{\dagger}+\hat{a})\hat{S}_z
\label{1}
\end{equation}
where $a$ is the annihilation operator for the field mode of
frequency  $\omega$, $\delta$ is the transition frequency of the
qubit, $\epsilon$ is the level asymmetry and $\lambda$ is the
strength of the coupling between the oscillator and the two-level
systems.

We assume a {\it slow} qubit and work in the regime
$\omega\gg\delta$ by employing the Born-Oppenheimer approximation,
\cite{liberti3,Sun}. The standard procedure is to separate the
Hamiltonian of Eq.(\ref{1}) in two parts, containing slow and fast
variables, respectively \cite{sainz}
\begin{equation}
   \hat{H}=\hat{H}_s+\hat{H}_f
    \label{pertplusunpert}
    \end{equation}
    where
    \begin{equation}
   \hat{H}_f=\omega \hat{a}^{\dagger}\hat{a} +\frac{\epsilon}{2} \hat{S}_z+
   \frac{\lambda}{\sqrt{N}}(\hat{a}^{\dagger}+\hat{a})\hat{S}_z,\quad \tilde{H}_s=-\frac{\delta}{2} \hat{S}_x
    \label{pertplusunpert0}
    \end{equation}
The eigenstates of the composite system can be written as a
coherent superposition of the eigenkets of $H_f$, having a
parametric dependence on (i.e. conditioned by) the values of the
slow variables:
\begin{equation}\label{deco}
    |\psi_{r}\rangle=\sum_{\{n,s_z\}}\phi_{s_z}^{(n,r)}|n\left[s_z\right]\rangle,\,(r=0,\dots,s)
\end{equation}
where the displaced number states of the oscillator are given by
\begin{equation}\label{Af}
|n\left[s_z\right]\rangle=
e^{-{\frac{\lambda}{\omega\sqrt{N}}}(a^\dagger-a)s_z}|n\rangle\otimes|s,s_z\rangle,
\end{equation}
They are the eigenstates of the fast Hamiltonian
\begin{equation}\label{Hf}
    \hat{H}_f|n\left[s_z\right]\rangle=V_n(s_z)|n\left[s_z\right]\rangle
    \, ,
\end{equation}
with eigenvalues
\begin{equation}\label{Vf}
    V_n(s_z)=\omega n+\frac{\epsilon}{2}
    s_z-{\frac{\lambda^2}{N\omega}}s_z^2\, .
\end{equation}
For different n, $V_n(s_z)$ contribute an effective adiabatic
potential felt by the slow subsystem so that the wave function
$\phi_{s_z}^{(n,r)}$ of the $N$ qubit system is determined by
\begin{equation}\label{Heff0}
\hat{H}_{eff}\phi_{s_z}^{(n,r)}=E_{(n,r)}(s_z)\phi_{s_z}^{(n,r)}
\, ,
\end{equation}
where the effective Hamiltonian is reduced to the form
\begin{equation}\label{Heff1}
    \hat{H}_{eff}=\frac{\delta}{2}
    \hat{S}_x+V_n(\hat{S}_z)=\omega \hat{a}^{\dagger}\hat{a}+
    \hat{H}_Z^{\perp,\,\parallel}\, ,
\end{equation}
with $\hat H_Z^{\perp, \, \parallel}$ being the Hamiltonian of the
uniaxial model introduced in the previous section, with coupling
constant $g_z= -{\frac{\lambda^2}{N\omega}}$:
\begin{equation}\label{uniaxial2}
    \hat{H}_Z^{\perp,\,\parallel}=-\frac{\delta}{2}
    \hat{S}_x+\frac{\epsilon}{2}
    \hat{S}_z+ g_z \hat{S}_z^2\, .
\end{equation}

The ground state of the coupled qubit-oscillator system is given
by
\begin{eqnarray}\label{deco3}
|\psi_{0}\rangle=\sum_{m=-N}^N\varphi_m
e^{-{\frac{m\lambda}{\sqrt{N\omega}}}(a^\dagger-a)}|0\rangle\otimes|N,m\rangle
\end{eqnarray}
where $\varphi_m\equiv\phi_m^{(0,0)}$.

The uniaxial (as well as the LGM model) and the Dicke model are
known to be equivalent in the thermodynamic limit. From the
discussion of this section, we see that there is a strict
relationship between the ground states of the two model-systems as
both can be expressed in the angular momentum basis with the same
amplitudes $\varphi_m$. However, this last equation shows that at
finite size there can be differences between their behaviors
since, in the case of the Dicke model these coefficients gets
effectively modified due to the presence of the displacement
operator, whose argument depends explicitly on the number of
qubits $N$. This implies that we will find small differences in
the $1/N$ expansions for the two models.

In order to continue the discussion on the Dicke model, we need to
evaluate the amplitudes $\varphi_m$. Therefore, we now turn our
attention to the uniaxial model with $N$ qubits. Once the
coefficients $\varphi_m$ are obtained, we will use them in Sec.
\ref{sect4} to complete the description of the finite size
behavior the Dicke model.

\section{Uniaxial model}\label{sect3}

\subsection{Continuum approach}

The Hamiltonian of a uniaxial model for a spin system with a
collective coupling can be written as
\begin{equation}\label{uniaxial}
    \hat{H}_Z^{\perp,\,\parallel}=-\frac{\delta}{2}
    \hat{S}_x+\frac{\epsilon}{2} \hat{S}_z-{\frac{g}{N}}\hat{S}_z^2
\end{equation}
with $\delta\geq0$ and where we have re-scaled the ferromagnetic
coupling constant by the number of spins, $g_z= -g/N$. This is
equivalent to the more diffusely found
$\hat{H}_X^{\perp,\,\parallel}$ Hamiltonian, that can be obtained
after the rotation $e^{i\pi S_y/4}$.

To connect this model to the discussion of the previous section,
one simply has to take $g=\lambda^2/\omega$.

The ground state of $\hat{H}_Z^{\perp,\,\parallel}$ lies in the
maximum spin sector $s\equiv N$. In this subspace, spanned by the
states $\{|N,m\rangle;\, m=-N,-N+2,\dots,N-2,N\}$, the ground
state can be written as
\begin{eqnarray}\label{deco2}
    |\phi_{0}\rangle=\sum_{m=-N}^N\varphi_m |N,m\rangle
\end{eqnarray}
where $\varphi_m$ are real coefficients.

We limit our present discussion to the case of the symmetric phase
$\epsilon=0$, that is the one relevant for the description of the
Dicke phase transition occurring at $\epsilon=0$ with $\lambda^2=
\omega \delta/4$, corresponding to $g= \delta/4$.

In the angular momentum basis, the Hamiltonian takes a
$(N+1)\otimes(N+1)$ tri-diagonal symmetric (Jacobi) matrix form
with double symmetries along both the main and the second
diagonal:
\begin{equation}\label{matr}
T_N=\left(
  \begin{array}{cccccccc}
    \lambda_{-N} & \delta_{-N} & 0 & \dots & 0 &0  &  0 \\
     \delta_{-N} & \lambda_{-N+2} &  \delta_{-N+2} & \dots &  0&0  & 0  \\
     0 & \delta_{-N+2} & \lambda_{-N+4} & \dots &0  &  0 &0\\
    \vdots & \vdots & \vdots & \ddots & \vdots & \vdots &  \vdots \\
    0 & 0 & 0 & \ldots & \lambda_{N-4} & \delta_{N-4} & 0\\
    0 & 0 & 0 & \ldots & \delta_{N-4} & \lambda_{N-2} & \delta_{N-2} \\
    0 & 0 & 0 & \ldots& 0 & \delta_{N-2} & \lambda_{N} \\
  \end{array}
\right)
\end{equation}
where
\begin{equation}\label{matrdel}
    \lambda_m=-\frac{\lambda^2}{N} m^2\,,\,-N\leq m\leq N
\end{equation}
and
\begin{equation}\label{matrndel}
   \delta_m=-\frac{\delta}{4}\sqrt{N(N+2)-m(m+ 2)} \, .
\end{equation}
These coefficients satisfy a confluence property
$\lambda_m/N\rightarrow\lambda(z)$ and
$\delta_m^2/N^2\rightarrow\delta^2(z)$  where $z=m/N$ as
$m,N\rightarrow\infty$ \cite{witte}. By application of theorems on
the zeros of orthogonal polynomials \cite{doorn} one finds that
the ground-state energy density in the $m,N\rightarrow\infty$
limit is given in general by
\begin{equation}\label{egs}
    \varepsilon_0(\infty)=\textrm{inf}\{\lambda(z)-2\delta(z)\}.
\end{equation}
Introducing the dimensionless parameter $\alpha=4g/\delta$ the
minimum is found at
\begin{equation}\label{min}
z_0=\left\{
 \begin{array}{ll}
 0, & \hbox{$(\alpha\leq1)$,} \\
 \pm \sqrt{1-1/\alpha^2}, & \hbox{$(\alpha>1)$,}
 \end{array}
\right.
\end{equation}
and the corresponding thermodynamic limit of the ground state
energy per spin is
\begin{equation}\label{egs2}
\lim _{N\rightarrow \infty} \frac{\varepsilon_0(N)}{N}=\left\{
 \begin{array}{ll}
 -\frac{\delta}{2}, & \hbox{$(\alpha\leq1)$,} \\
 -\frac{\delta}{4}\left(\alpha+\frac{1}{\alpha}\right), & \hbox{$(\alpha>1)$.}
 \end{array}
\right.
\end{equation}

For finite $N$, the solution of the eigenvalues problem for the
ground state reduces to the recurrence relation
\begin{equation}\label{rec}
\delta_{m-2}\varphi_{m-2}+\lambda_m\varphi_m+
\delta_m\varphi_{m+2}=\varepsilon_0\varphi_m
\end{equation}
that can be rewritten as a second order linear difference equation
\begin{eqnarray}\label{rec2}
  && 2(\delta_m+\delta_{m-2})\bigtriangleup_2\varphi_{m}+
2(\delta_m-\delta_{m-2})\bigtriangleup_1\varphi_{m}\nonumber\\
&+&(\delta_m+\delta_{m-2}+\lambda_m)\varphi_m=\varepsilon_0\varphi_m
\end{eqnarray}
where
$\bigtriangleup_2\varphi_{m}=(\varphi_{m+2}+\varphi_{m-2}-2\varphi_{m})/4$
and $\bigtriangleup_1\varphi_{m}=(\varphi_{m+2}-\varphi_{m-2})/4$
are finite differences of second and first order, respectively.

A simple analytic behavior of the coefficients $\varphi_m$ for
$N\gg1$ can be derived by considering $m/N$ as a continuous
variable, and by expanding the recursion relation (\ref{rec2}) in
series around the minima of Eq. (\ref{egs}). For $\alpha\leq1$,
expanding in series (\ref{rec}) around $m=0$ and neglecting
corrections of order $1/N^2$, one obtains
\begin{equation}\label{phim0}
    \varphi_m^{''}+\left[\frac{\varepsilon_0(N)}{N\delta}+\frac{1}{2}\left(1+\frac{1}{N}\right)-
\frac{1-\alpha}{4N^2}m^2\right]\varphi_m\simeq 0
\end{equation}
whose solution is
\begin{equation}\label{phim}
    \varphi_m\simeq\left(\frac{2k}{\pi N}\right)^{1/4}e^{-km^2/4N}
\end{equation}
with $k=\sqrt{1-\alpha}$. The ground state energy per spin is given by
\begin{equation}\label{en1}
    \frac{\varepsilon_0(N)}{N}\simeq-\frac{\delta}{2}\left(1+\frac{1-\sqrt{1-\alpha}}{N}\,\right)\,.
\end{equation}

For $\alpha>1$, by expanding in series  Eq. (\ref{rec}) around
$m\simeq\pm m_0=\pm N\sqrt{1-1/\alpha^2}$, one gets
\begin{eqnarray}\label{phim00}
    \varphi_m^{''}&+&\alpha\left[\frac{\varepsilon_0(N)}{N\delta}+\frac{1}{4}\left(\alpha+\frac{1}{\alpha}\right)\right.
\nonumber\\
&+&\left.\frac{\alpha}{2N}-\frac{\alpha(\alpha^2-1)}{4N^2}(m\pm m_0)^2\right]\varphi_m\simeq0
\end{eqnarray}
whose approximate solution is the symmetric superposition
\begin{equation}\label{phim2}
    \varphi_m\simeq\frac{1}{\sqrt{2}}\left(\varphi_m^++\varphi_m^-\right)
\end{equation}
with
\begin{equation}\label{phim3}
    \varphi_m^{\pm}=\left(\frac{2\bar{k}}{\pi N}\right)^{1/4}e^{-\bar{k}(m\mp m_0)^2/4N}
\end{equation}
where $\bar{k}=\alpha\sqrt{\alpha^2-1}$. In this regime one has
\begin{equation}\label{en2}
    \frac{\varepsilon_0(N)}{N}\simeq-\frac{\delta}{2}\left[\frac{1}{2}\left(\alpha+\frac{1}{\alpha}\right)+
\frac{\alpha-\sqrt{\alpha^2-1}}{N}\right]\,.
\end{equation}

In this language, the transition is  readily understood:  above
the coupling value corresponding to $\alpha=1$ a drastic change in
the form of the ground state wave function takes place, with a
breaking of the ``inversion'' symmetry around $m=0$. For a finite
size system, the transition becomes smother and smother and the
wave function $\varphi_m$ gradually changes from a one peaked
gaussian to the superposition with two peaks  that emerge
progressively as the value of $z_0$ moves away from the origin
(i.e., as $\alpha$ increases).

\begin{figure}
\includegraphics[width=8.5cm]{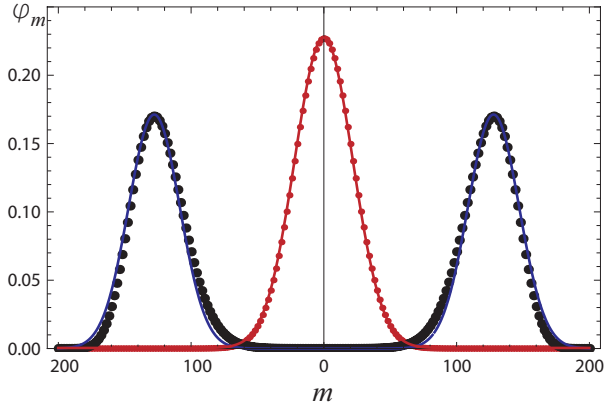}\\
\caption{\label{wf0} (Color online): Normalized $\varphi_m$
function for $\alpha=0.3$ (small red circles) and $\alpha=1.3$
(black circles) for a spin system of size $N=200$. Comparison is
made with the analytic expressions of Eqs.(\ref{phim}-\ref{phim2})
(continuous lines).}
\end{figure}

For large enough $N$, we can check the continuum approximation by
comparing it to the behavior obtained by solving the tridiagonal
matrix numerically. In Fig.(\ref{wf0}) $\varphi_m$ for $N=200$ is
shown with $\alpha=0.3$ and $\alpha=1.3$ compared with the
analytic expressions of Eqs.(\ref{phim}-\ref{phim2}).

\subsection{Finite-size corrections}

Having obtained the ground state coefficients $\varphi_m$
(together with the ground state energy), we may evaluate the
average values of every physical observable; in particular, we
concentrate on the total spin components. One immediately gets
\begin{eqnarray}\label{sxsysz}
  \frac{\langle
   S_x\rangle}{N}&=&
    -\frac{2}{N}\frac{\partial\varepsilon_{0}(N)}{\partial \delta}\nonumber\\
&=& \left\{
 \begin{array}{ll}
 1+\frac{1}{N}\left(1+\frac{\alpha-2}{2\sqrt{1-\alpha}}\right), & \hbox{$(\alpha\leq1)$;} \\
 \frac{1}{\alpha}+\frac{1}{N\sqrt{\alpha^2-1}}, & \hbox{$(\alpha>1)$.}
 \end{array}
\right.
\end{eqnarray}
and
\begin{eqnarray}\label{2sz}
   \frac{\langle
   S_z^2\rangle}{N^2}&=& -\frac{1}{N}\frac{\partial\varepsilon_{0}(N)}{\partial g}\nonumber\\
&=& \left\{
 \begin{array}{ll}
 \frac{1}{N\sqrt{1-\alpha}}, & \hbox{$(\alpha\leq1)$;} \\
 1-\frac{1}{\alpha^2}+\frac{2}{N}\left(1-\frac{\alpha}{\sqrt{\alpha^2-1}}\right), & \hbox{$(\alpha>1)$.}
 \end{array}
\right.
\end{eqnarray}
The  expression for $\langle S_{x,y}^2\rangle$ are, instead, a bit
more complicated
\begin{eqnarray}\label{2sx}
  {\langle
   S_{x,y}^2\rangle}&=&\frac{1}{2}\left[{N(N+2)}-{\langle
   S_z^2\rangle}\right]\nonumber\\
   &\pm&2\sum_{m=-N+2}^{N-2}a_m^+a_{m}^-\varphi_{m-2}\varphi_{m+2}
   \, .
\end{eqnarray}
However, they can be simplified by making use of the simple
results $\varphi_{m-2}\varphi_{m+2}=e^{-2k/N}\varphi_{m}^2$  for
$\alpha\leq1$ and
$\varphi_{m-2}\varphi_{m+2}=e^{-2\bar{k}/N}\varphi_{m}^2$ for
$\alpha>1$ that are easily derived from our analytic expressions
for $\varphi_m$. Thus, one obtains
\begin{equation}\label{2sxlim}
  \frac{\langle
   S_x^2\rangle}{N^2}\simeq \left\{
 \begin{array}{ll}
 1+\frac{2}{N}\left(1-\frac{1}{\sqrt{1-\alpha}}\right), & \hbox{$\alpha\leq1$;} \\
\frac{1}{\alpha^2}+\frac{1}{N}\frac{\alpha^2+1}{\alpha\sqrt{\alpha^2-1}}, & \hbox{$\alpha>1$.}
 \end{array}
\right.
\end{equation}
\begin{equation}\label{2sylim}
  \frac{\langle
   S_y^2\rangle}{N}\simeq \left\{
 \begin{array}{ll}
 {\sqrt{1-\alpha}}, & \hbox{$\alpha\leq1$;} \\
{\sqrt{1-\frac{1}{\alpha^2}}}, & \hbox{$\alpha>1$.}
 \end{array}
\right.
\end{equation}

In the region $\alpha\sim1$ we must take into account also the
next to leading order in the expansion of the recursion relation
(\ref{rec}) that gives a non negligible contribution near the
phase transition point. We thus need to consider the
quartic-oscillator-like equation
\begin{equation}\label{phimcrit}
    \varphi_m^{''}+\left[\frac{\varepsilon_0(N)}{N\delta}+\frac{1}{2}\left(1+\frac{1}{N}\right)-
\frac{1-\alpha}{4N^2}m^2-\frac{m^4}{16N^4}\right]\varphi_m\simeq 0
\, .
\end{equation}
Using the approach presented in a previous work \cite{liberti3},
the equation (\ref{phimcrit}) can be reduced to a
single-parametric problem with the help of Symanzik scaling
procedure \cite{simon}. This is done, by re-casting the equation
(\ref{phimcrit}) into the equivalent form
\begin{equation}\label{hamscaling}
 \varphi_n^{''}+\left(e_{0}\left(\zeta\right)-\zeta
n^2-n^4\right)\varphi_n\simeq0
\end{equation}
where $n=m(2N)^{-2/3}$ is a scaled variable. The only
remaining scale parameter is then
$\zeta=(2N)^{2/3}(1-\alpha)$, while the
ground-state energy is rewritten as
\begin{equation}\label{sr}
\frac{\varepsilon_{0}(N)}{N}=
-\frac{\delta}{2}\left(1+\frac{1}{N}\right)+\delta\frac{e_{0}(\zeta)}{(2N)^{4/3}}
\end{equation}
For $\zeta \sim 0$ (that is, very close to the transition point),
we can resort to perturbation theory and obtain the ground state
energy as an expansion in powers of $\zeta$,
\begin{equation}\label{ps}
e_{0}(\zeta)=\sum_{n=0}^\infty\beta_n\zeta^n\, .
\end{equation}
It is easy to show that $\beta_0= e_{0}(0)\simeq 1.06036$ is the
lowest  eigenvalue of the pure quartic oscillator and
$\beta_1=e_0^\prime(0)\simeq 0.36203$.

Using these results to obtain an approximate expression for the
ground state energy and for the coefficients $\varphi_m$, it is
easy to derive the following leading nontrivial finite-size
corrections for one- and two-spin correlation functions

\begin{equation}\label{sxcrit}
     \frac{\langle
   S_x\rangle}{N}\simeq1-\frac{2\beta_1}{(2N)^{2/3}}
\end{equation}
\begin{equation}\label{sz2crit}
     \frac{\langle
   S_z^2\rangle}{N^2}\simeq\frac{4\beta_1}{(2N)^{2/3}}
\end{equation}\begin{equation}\label{sx2crit}
     \frac{\langle
   S_x^2\rangle}{N^2}\simeq1-\frac{4\beta_1}{(2N)^{2/3}}
\end{equation}\begin{equation}\label{sy2crit}
     \frac{\langle
   S_y^2\rangle}{N^2}\simeq\frac{8\beta_0}{3(2N)^{4/3}}
\end{equation}
The critical exponents in these expressions are in full agreement
with those reported in Ref.\cite{dusuel}. Our method allowed us to
obtain also the pre-factors, that cannot be determined with
typical scaling arguments and that are important to transfer these
results to the case of the Dicke model.

\begin{figure}
 \includegraphics[width=8.5cm]{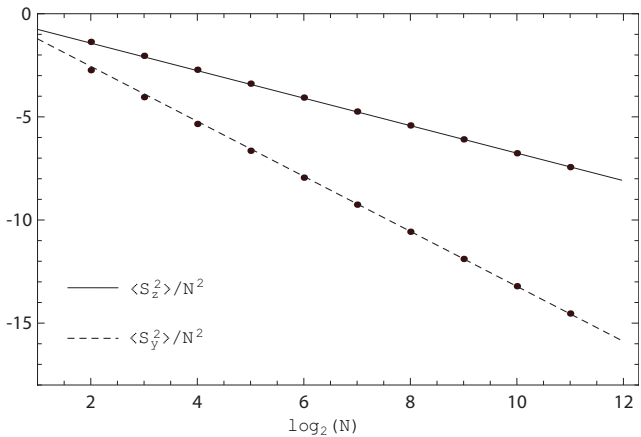}\\
 \caption{\label{cs} Scaling of two-spins correlation functions ${\langle
S_z^2\rangle}/{N^2}$ and ${\langle
 S_y^2\rangle}/{N^2}$ as a function of $N$
($\textrm{log}-\textrm{log}$ plot) at the critical point
$\alpha=1$.}
\end{figure}

In Fig.(\ref{cs}) we make a comparison of the analytical results
for the leading nontrivial finite-size corrections with those
obtained from a direct numerical solution at the critical point.
One can see that the agreement is good even for small values of
$N$.

\subsection{Ground state Entanglement}
Before going back to the Dicke model, we use the results we have
obtained in order to discuss the critical behavior of the ground
state entanglement for the uniaxial model. In this respect, it is
useful to make a partition of the $N$ spins in two blocks of size
$L$ and ($N-L$), respectively. Using the decomposition
\begin{equation}\label{bip}
    |N,m\rangle=\sum_{l=-L}^L
    p_{lm}^{1/2}|N-L,m-l\rangle\otimes|L,l\rangle
\end{equation}
where
\begin{equation}\label{perm}
    p_{lm}=\frac{\left(%
\begin{array}{c}
  L \\
  \frac{L+l}{2} \\
\end{array}%
\right)\left(%
\begin{array}{c}
  N-L \\
  \frac{N-L+m-l}{2} \\
\end{array}%
\right)}{\left(%
\begin{array}{c}
  N \\
  \frac{N+m}{2} \\
\end{array}%
\right)} \,
\end{equation}
we obtain the ground-state reduced density matrix of the block of
size $L$ out of the total $N$ spins in the form
\begin{eqnarray}\label{tofv2}
    \rho_{L,N}&=&\sum_{l_1=-L}^L\sum_{l_2=-L}^L|L,l_1\rangle\langle
    L,l_2|\nonumber\\
    &\times&\sum_{m=-N}^Np_{l_1m}^{1/2}p_{l_2m-l_1+l_2}^{1/2}\varphi_{m}\varphi_{m-l_1+l_2}
    \, .
\end{eqnarray}
We then compute the linear entropy as a measure of the
entanglement of the block of size $L$ with the rest of the system.

\begin{equation}
\tau_{L} = \eta_L \Bigl [1 -\mbox{Tr}\left( \rho_{L,N}^2
\right)\Bigr ]
\end{equation}
where the pre-factor is chosen to be $\eta_L = \frac{2^L}{2^L-1}$
in order to bound $\tau_L$ to $1$.

In particular, for $L=1$, the state of every single qubit is found
to be
\begin{equation}\label{rho1N}
\rho_{1,N}=\frac{1}{2}\left(I+\frac{\langle
S_x\rangle}{N}\sigma_x\right)
\end{equation} where $I$ is the identity. We can then evaluate the one-tangle as
\begin{equation}\label{tau2}
\tau_1= 2 \Bigl [1 -\mbox{Tr}\left( \rho_{1,N}^2 \right)\Bigr ]
\equiv 1-\frac{\langle
S_x\rangle^2}{N^2} \, .
\end{equation}
One has
\begin{equation}\label{tau2lim}
  \tau_1 \simeq \left\{
 \begin{array}{ll}
 \frac{1}{N}\left(2+\frac{\alpha-2}{\sqrt{1-\alpha}}\right), & \hbox{$\alpha\leq1$;} \\
 1-\frac{1}{\alpha^2}+\frac{2}{N\alpha\sqrt{\alpha^2-1}}, & \hbox{$\alpha>1$.}
 \end{array}
\right.
\end{equation}
and
\begin{equation}\label{tau1crit}
    \tau_1 \simeq \frac{4\beta_1}{(2N)^{2/3}}\,,\quad \alpha=1
\end{equation}
The reduced density matrix of two qubits ($L=2$), can be written
in the angular momentum basis $\{|2,m\rangle \,\}$, with
$m=2,0,-2$. In general, one should also consider the state
$|0,0\rangle$; but its population is zero in our case, so that we
can erase the corresponding line and row, and write $\rho_{2,N}$
in the form:
\begin{equation}\label{rhoi}
    \rho_{2,N}=\left(%
\begin{array}{cccc}
  v_+ & \sqrt{2} x_+ & u \\
  \sqrt{2} x_+ & 2 w & \sqrt{2} x_- \\
  u & \sqrt{2} x_- & v_- \\
\end{array}%
\right)
\end{equation}
where the matrix elements may be expressed in terms of the
expectation values of the collective operators as \cite{wang}
\begin{eqnarray}
  v_{\pm} &=& \frac{N(N-2)+\langle S_z^2\rangle}{4N(N-1)}\pm\frac{\langle S_z\rangle}{2N}\\
  w &=& \frac{N^2-\langle S_z^2\rangle}{4N(N-1)}\\
  u &=& \frac{\langle S_+^2\rangle}{N(N-1)} \\
  x_\pm &=& \frac{\langle S_+\rangle}{2N}\pm\frac{\langle [S_+,S_z]_+\rangle}{4N(N-1)}
\end{eqnarray}

The entanglement between two qubits can be expressed in terms of
the concurrence \cite{Wootters}. Since the ground state lies in
the maximum spin sector and has real coefficients in the basis
$\{|N,m\rangle\}$, one has
\begin{equation}\label{conc}
    C=\textrm{max}\{0,C_y\}
\end{equation}
where
\begin{equation}\label{cy}
   (N-1) C_y=1-\frac{\langle S_y^2\rangle}{N}.
\end{equation}
Thus, the concurrence needs to be re-scaled by the factor $N-1$
(thus, $C$ vanishes $\sim 1/N$ in the thermodynamic limit), with
$C_r = (N-1) C$. In the thermodynamic limit, only $C_r$ remains
finite:
\begin{equation}\label{conclim}
  C_r\simeq \left\{
 \begin{array}{ll}
 1-\sqrt{1-\alpha}, & \hbox{$\alpha\leq1$;} \\
1-\sqrt{1-\frac{1}{\alpha^2}}, & \hbox{$\alpha>1$.}
 \end{array}
\right.
\end{equation}
For finite $N$ and at the critical point, Eq.(\ref{sy2crit}) gives
\begin{equation}\label{cycrit}
    C_r\simeq1-\frac{4\beta_0}{3\,(2N)^{1/3}} \, .
\end{equation}
This shows that, at the critical point, the behavior of the
concurrence is modified and $C_r$ scales with $N$ with a critical
exponent of $1/3$.

\section{Finite size scaling of the Dicke model}
\label{sect4}
\begin{figure}
 \includegraphics[width=8.5cm]{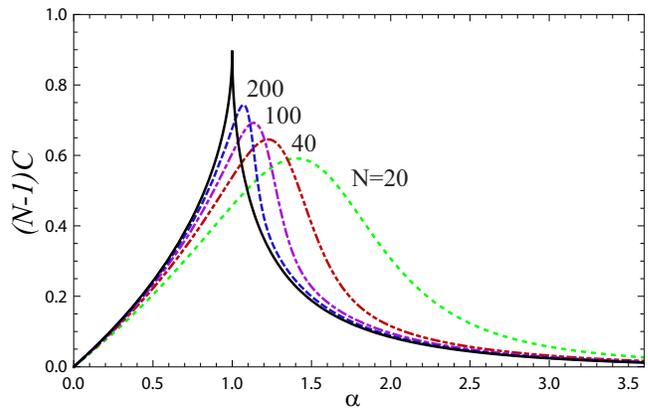}\\
\caption{\label{concd} (Color online): Scaled concurrence for the
Dicke model as a function of $\alpha$, for $D=0.1$ and for system
sizes $N=10, 20, 40, 100$ and $\infty$ (bottom to top).}
\end{figure}
We can now make use of the expressions of the amplitudes
$\varphi_m$  obtained for uniaxial model in order to discuss the
finite size behavior of the Dicke model.

Whenever we are interested in a qubit observable, that is,
whenever the result can be obtained by tracing out the oscillator,
the only difference with the uniaxial model is the appearance of
exponentials of the form $\exp\{-\lambda^2 (m-m')^2/N \omega^2
\}$, due to the overlap of different coherent states. This kind of
terms modifies the behavior of the spin observables for small $N$,
but  for very large $N$ one can expect to obtain very similar
behaviors for the Dicke and the uniaxial models. One finds
\begin{equation}\label{sxmod}
\langle
   S_x\rangle\rightarrow e^{-\frac{\alpha D}{2N}}\langle
   S_x\rangle
\end{equation}
and
\begin{equation}\label{sxsymod}
\langle
   S_x^2-S_y^2\rangle\rightarrow e^{-\frac{2\alpha D}{N}}\langle
   S_x^2-S_y^2\rangle
\end{equation}
which coincides, respectively, with Eq. (\ref{sxsysz}) and Eq. (\ref{2sx}) in the $D=\delta/\omega\rightarrow 0$ limit.
Using these results, one can show, for example, that the ground state energy at the critical point reads
\begin{equation}\label{srd}
\frac{\varepsilon_{0}(N)}{N}\simeq
-\frac{\delta}{2}\left(1+\frac{2-D}{2N}-\frac{2\beta_{0}}{(2N)^{4/3}}\right).
\end{equation}

Once all of the average values of the spin observables are
obtained, it is easy to get expressions for the various
entanglement measures. In particular, the re-scaled concurrence in
the thermodynamic limit, reads
\begin{equation}\label{conclimd}
   C_r\simeq \left\{
 \begin{array}{ll}
 1-D\alpha-\sqrt{1-\alpha}, & \hbox{$\alpha\leq1$;} \\
1-\frac{D}{\alpha}-\sqrt{1-\frac{1}{\alpha^2}}, & \hbox{$1<\alpha<\alpha_0$.}
 \end{array}
\right.
\end{equation}
where $\alpha_0=(1+D^2)/2D=(\delta^2+\omega^2)/2\omega\delta$. For
finite size, at the critical point one gets
\begin{equation}\label{cycritd}
    C_r\simeq1-D-\frac{4\beta_0}{3\,(2N)^{1/3}}
\end{equation}
Thus, the concurrence scales with $N$ exactly as in the uniaxial
model. Fig. (\ref{concd}) shows the $C_r$ both for finite $N$ and
for $N \rightarrow \infty$.

The difference between  the adiabatic Dicke model and the uniaxial
one lies in the presence of the oscillator, which is far detuned
from the spins but that can still be excited (because of the
presence of the counter-rotating terms in the Hamiltonian) and
becomes correlated with the qubits. In particular, the
entanglement between the oscillator and the $N$ qubits, can be
evaluated by  the linear entropy which is of the form
\begin{equation}\label{line}
\tau_N = \eta_N \Bigl ( 1- \mbox{Tr} \left \{ \rho_{N}^2 \right \}
\Bigr ) \, ,
\end{equation}
where $\rho_{N}$ is the reduced density matrix for the N-qubits
sub-system, obtained from the ground state density operator
(\ref{deco3}) by tracing out the field variables
\begin{eqnarray}\label{tofvd}
    \rho_N&=&\textrm{Tr}_F\{|\psi_{0}\rangle\langle\psi_{0}|\} \\
    &=&\sum_{m_1,m_2=-N}^Ne^{-\frac{\alpha D}{8N}(m_1-m_2)^2}\varphi_{m_1}\varphi_{m_2}|N,m_1\rangle\langle
    N,m_2|\nonumber
\end{eqnarray}
Evaluating the trace of $\rho_N$ squared, one has
\begin{equation}\label{line2}
\tau_N = \eta\left( 1- \sum_{m_1,m_2=-N}^Ne^{-\frac{\alpha D}{4N}{{(m_1-m_2)}^2}}\varphi_{m_1}^2\varphi_{m_2}^2
\right)
\end{equation}

\begin{figure}
 \includegraphics[width=8.5cm]{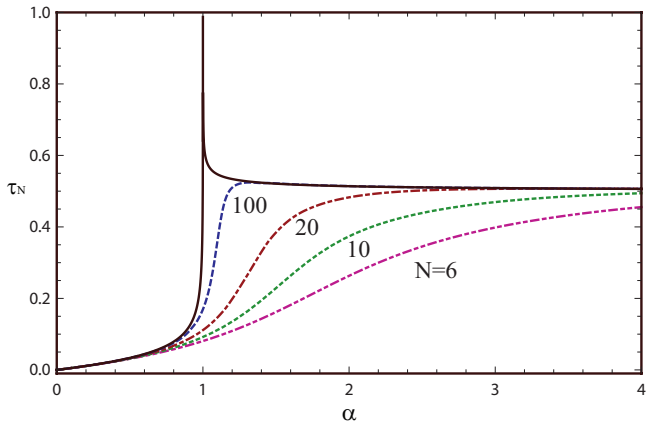}\\
 \caption{\label{tanglefiN} (Color online): The tangle $\tau_N$ between the oscillator
 and the $N$ qubits as a function of $\alpha$, for $D=0.1$ and
for system sizes $N=6, 10, 20, 100$ and $\infty$ (bottom to top).}
\end{figure}

In the thermodynamic limit the sum can be computed exactly to get
\begin{equation}
\tau_{\infty} =\left\{%
\begin{array}{ll}
   1- \left (1+ \frac{D\alpha}{\sqrt{1-\alpha}} \right )^{-\frac{1}{2}} & \hbox{$(\alpha\leq 1)$} \\
   1-\frac{1}{2} \left (1+ \frac{D}{\sqrt{\alpha^2-1}} \right )^{-\frac{1}{2}} & \hbox{$(\alpha> 1),$} \\
\end{array}%
\right. \, ,
\end{equation}
which shows a cusp at the critical point, where $\tau_{\infty}
=1$. Fig. (\ref{tanglefiN}) shows $\tau_N$ both for finite $N$ and for
$N \rightarrow \infty$.
When $N$ is very large ($N\gg 4/D^3$), the entanglement scales as
\begin{equation}\label{tNsa}
\tau_N(\alpha=1)  \sim 1-  \,
K\left({\frac{\pi}{D}}\right)^{1/2}\left({\frac{4}{N}}\right)^{1/6} \, ,
\end{equation} where $K= \frac{1}{4}\int dn \varphi_n^4
\simeq 0.46$, and $\varphi_n$ is the normalized solution of Eq.
(\ref{hamscaling}) for $\alpha=1$. The fact that the leading term
in the $1/N$ expansion of $\tau_N$ has exponent $1/6$, implies
that the convergence of the series is slower (with respect to
those found for other physical quantities) and that for small
values of $N$, subsequent terms should be taken into account.
\section{Concluding remarks}\label{sect5}
We have discussed the finite size critical behavior of the Dicke
model for the case of a fast oscillator coupled to many slower
qubits. We have derived a direct relationship between this system
and the uniaxial model, describing the collective interaction
among qubits residing on a fully connected graph. In particular,
we have obtained a precise one-to-one correspondence between the
energy eigen-states of the two models, showing that their critical
behavior are closely related both in the thermodynamic limit and
at finite size. We have then, adopted a continuum approximation in
order to describe analytically the ground state of the uniaxial
model which we used to re-obtain all the known features of the
model, such as its critical exponents. We have also obtained a
full characterization of the $1/N$ expansion (including non
universal features such as the pre-factors) for many physical
observables, among which we dedicated a particular emphasis to the
description of the entanglement content of the ground state and to
its critical behavior.

Using the solution obtained for the uniaxial model, we were then
able to go back to the original Dicke model and to describe its
critical behavior and its scaling properties, again obtaining not
only the scaling exponents for various physical quantities, but
their entire $1/N$ expansions (of which the first terms are shown
and discussed explicitly).

The two models we have discussed obviously differ because of the
presence of the bosonic mode in the Dicke case. From a physical
point of view this implies that an entanglement is built up  not
only among the qubits (as in the uniaxial model) but also between
qubits and oscillator. Formally, this manifests itself in the fact
that the oscillator state is a displaced vacuum (i.e. coherent)
state conditioned on the qubit magnetization in the direction of
the coupling (i.e., on the value of $S_z$, in our notation). The
presence of these quantum correlation with the oscillator also
modifies the entanglement among qubits (formally, because of the
presence of some exponential pre-factors that essentially suppress
entanglement), and this can be interpreted in terms of the
monogamy of entanglement.

Apart from this, the two models have many features in common; in
particular, their critical behavior are closely related and their
quantum phase transition are essentially the same, occurring at
the same point in parameter space (once the proper relationship
between the physical parameters is taken into account).

%

\end{document}